\documentclass[letter,11pt]{article}

\usepackage{float}
\usepackage{graphics,graphicx}
\usepackage{hyperref,setspace}
\usepackage{epstopdf}

\begin{document}

\begin{center}
\includegraphics[width=\textwidth]{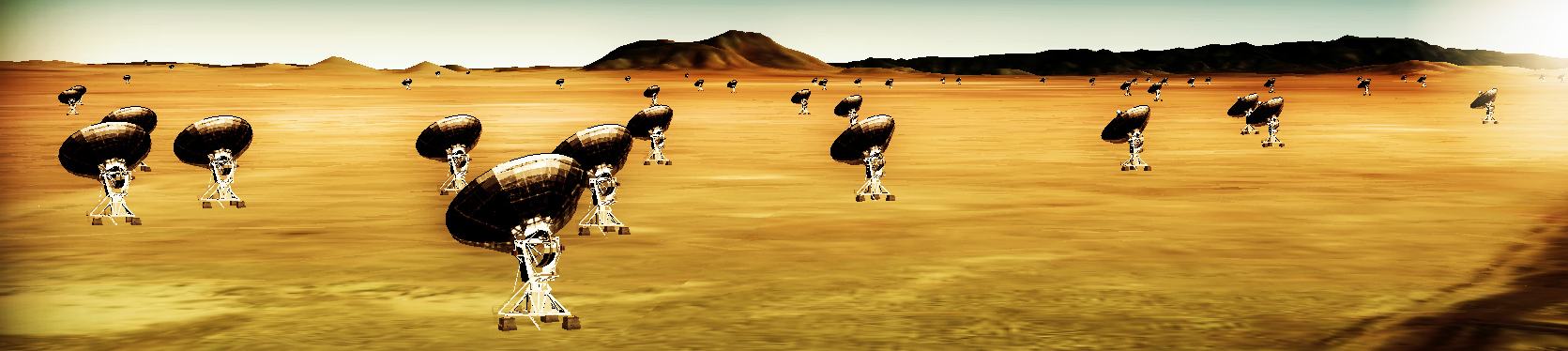}
\end{center}

\begin{center}

{\bf \large Next Generation Very Large Array Memo No. 78}

\vspace{0.1in}

{\bf \large High Resolution, Wide Field, Narrow Band, Snapshot Imaging}

\end{center}

\hrule 

\vspace{0.3cm}

\centerline{C.L. Carilli, E. Murphy, V. Rosero (NRAO), K. Mooley (NRAO, Caltech)}
\centerline{E. Jim\'enez-Andrade, K. Golap, B. Butler (NRAO)}

\vspace{0.3cm}


\begin{abstract}

We investigate the imaging performance of an interferometric array in the case of wide field, high resolution, narrow band, snapshot imaging. We find that, when uv-cell sizes are sufficiently small (ie. image sizes are sufficiently large), each instantaneous visibility record is gridded into its own uv-cell. This holds even for dense arrays, like the core of the ngVLA. In this particular, application, Uniform weighting of the gridded visibilities approaches Natural weighting, with its often deleterious consequences on the resulting synthesized beam. For a core-dominated array, we show that the resulting image noise is highly correlated on scales comparable to the spatial frequencies of the core baselines. In general, this study accentuates the fact that, for imaging applications that require high resolution (Plains array and greater), many of the core antennas can be employed as a separate subarray for low resolution science, without sacrificing the quality of the high resolution science.

\end{abstract}

\section{Introduction}

Rapid, wide field, high-resolution imaging in radio astronomy has taken on new relevance, with the discovery of millisec events, such as fast radio bursts, and the possibility of detecting prompt emission from gravitational wave sources. The need for arc-second localization is crucial to identify the parent object of the transient event. Likewise, very wide field surveys require on-the-fly mapping, implying short integrations on a given position. These new science goals have led to an increasing use of short, 'snapshot' observations, corresponding in many cases to a single time-record, down to 10msec in the case of FRB searches (Law et al. 2018a,b), from which to synthesize the sky image. 

In this study, we consider the imaging capabilities of a radio synthesis array in the context of instantaneous snapshot (single time-record), narrow band observations. We employ the latest configuration of the Next Generation Very Large Array (ngVLA), but also consider the current VLA as well. Besides the implications of a short integration, the ngVLA has a dense core, with roughly half the total collecting area on baselines less than 1.3~km, with the rest extending to long baselines, out to tens of km, or more. Similar core-dominated arrays are being designed at meter through centimeter wavelengths, including SKA-low and SKA-mid. Such arrays, with their large spatial dynamic range weighted heavily toward short baselines, will display more pronounced differences due to gridding of the short vs. long baselines, relative to arrays without a dense core, such as the VLA and ALMA.

We consider only thermal noise (ie. no calibration errors) and investigate the effect on the synthesized beam shape and image noise, of the weighting of the visibilities, the length of the synthesis observation, and the field size. Typically, in synthesis imaging, the terms that are thought to dominate the image results are the uv-weighting and the pixel size. However, we find for narrow band, instantaneous, snapshot observations, the field size becomes a critical parameter as well, through the uv-cell size generated in the imaging/gridding process. 

We should point out at the start that narrow band, snapshot, wide field, high resolution observations are not likely going to be common, for a few reasons. For wide field continuum imaging, bandwidth synthesis will be employed, which alters the weighting of inner and outer baselines via the gridding and uv-weighting process. For wide field spectral line OTF mosaics, bandwidth synthesis is not possible, but such observations are typically done at low spatial resolution. For many fast transients, yes, the pulse signal is dispersed in frequency, ie. later arrival times at low frequencies, due to the frequency dependence of the index of refraction of the interstellar medium. However, the de-dispersion is typically done in the visibility domain, and then bandwidth synthesis is again employed (Law et al. 2018a,b). More directly relevant are the rare, narrow band, less-dispersed pulses from magnetic flare stars, or white dwarf pulses.  Moreover, the study below uncovers some interesting implications of interferometric gridding and weighting, in particular in the context of core-dominated arrays. While we have investigated the extremes of the process, the noted effects due to uv-grid cell size will be manifest, although less pronounced, even in non-snapshot, wider band observations.

\section{Sky and Telescope Models, and Simulations}

We employ the ngVLA Rev C configuration including the Plains + Core arrays (Selina et al. 2017). The array has 168 antennas, including 94 antennas in the Core to 1.3km maximum baselines, and the rest extending symmetrically from the core in five spiral arms, with maximum baselines to 37km. This array is chosen to provide adequate resolution (arcsecond) to localize sources, but not require excessively large images that would challenge post-processing. The antenna layout and uv-coverage are shown in Figure 1, including the full configuration, and a blow-up of the core.

For the snapshot observation, we assume the current minimum integration time for ngVLA Rev C of 100~msec, as might be employed in the initial wide field searches for fast transients with the ngVLA. For comparison, we also perform a 3000s synthesis simulation, using 10s records, and we include similar observations using the VLA A configuration. All of the simulations are done at 2.4 GHz. 

The sky and telescope model are folded through the SIMOBSERVE process as described in Carilli et al. (2017).  We then insert noise per visibility using the setnoise tool in CASA. We adopt a noise value based on the Rev C system parameters, assuming channel width of 4MHz, corresponding to the channel widths proposed for fast, dispersed transient searches with the ngVLA by Law et al. (2019b). The resulting noise in the 100~msec images made using Natural weighting is 2.0 mJy/beam (Selina et al. 2018). 

\section{Imaging}

We employ the CASA TCLEAN algorithm for the imaging. We adopt a $0.2"$ pixel size in all cases,  For reference, the maximum baseline in the array is 37 km, or $B_{max}$ = 310 k$\lambda$. To Nyquist sample this visibility requires a pixel size = $0.33"$ (= 1/(2$B_{max}$).  

We then adjusted the weighting to either pure Natural (NA) or Uniform (UN). For reference, NA weighting implies all visibilities get the same weight, such that uv-grid cells with many visibility records are given high weight in the Fourier transform to the image domain.  UN weighting implies all gridded cells get the same weight, such that multiple visibilities that are gridded into a single cell get down-weighted by the number of visibilities in the cell (Briggs, Schwab, Sramek 1999). 

We make images of three different sizes. The largest field size is $50'$, corresponding to the full field of view to the 10\% power level of the primary beam at 2.4 GHz of the ngVLA 18m antenna. We also make fields of $5'$ and $0.5'$, for comparison. An important caveat is that we do not use facet imaging to make the wide field. Breaking the field into smaller facets will increase the grid cell size, and alter the results for different uv-weighting schemes. We return to this point below.

Our primary analysis entails just noise visibility data sets, ie. no cosmic sources. We do not perform any deconvolution of these noise fields. However, we include one image with a point source to demonstrate the effect on the resulting source size, and deconvolve this image to a minimum residual clean component of $2\sigma$.

\section{Effect of Field Size and Rotation Synthesis}

The field size is relevant for the following reason. The UV-cell size is set by the inverse of twice the image size. For an image size of $\theta_i = 3000" = 0.0145$rad, and wavelength of $\lambda = 0.125$~m, this implies a uv-cell size = 1/($2\theta_i$) = 34 $\lambda$ = 4.3~m. For comparison, the shortest baseline in the core of the ngVLA is 30m. For the $300"$ field, the uv-cell size increases to 43~m (340 $\lambda$), and for the $30"$ field, the cell size is 430~m (3400 $\lambda$). 

The implication is that for the large field, almost all visibilities in an instantaneous snapshot observation fall into their own, individual uv-grid cell, with very few 'crossing points' coming from different baselines that happen to project to the same uv-cell. Hence, UN weighting will approach NA weighting, since each cell only contains one visibility, even for the densely packed core baselines. 

For the smaller images, the uv-grid cells are successively larger, such that each cell contains many visibilites. In the case of the smallest, $30"$, field size, all the core baselines fall into just a handful of uv-grid cells. Hence, there will be a dramatic difference between NA and UN weighting.

The instantaneous snapshot coverage of the uv-plane is shown in Figure 2. The coverage of the full configuration is shown, plus blow-ups of the core visibilities (out to $\sim 1.3$km baselines), and a further blow-up of a dense region in the core uv-plane. In the latter, we include two boxes, corresponding to the uv-grid cell size for the $3000"$ field and the $300"$ field. Clearly, the large field affords about one grid cell per visibility. The smaller field leads to many visibilities falling into a given uv-cell. The $30"$ field uv-cell size would be much larger than the entire region shown in Figure 2c.

An interesting comparison is to allow for a 3000s synthesis observation, with 10s records. The resulting uv-plane coverage is shown in Figure 3. Shown is the full uv-plane, and blow-ups of an inner and outer region of the uv-plane, centered at uv-distances of 160~m (1.28k$\lambda$) and 14~km (112~k$\lambda$), respectively. Also shown is the uv-cell size for the $50'$ field, as per figure 2. The uv-tracks rotate through the uv-plane, and would make a full circle in 24hr, so the tracking distance, $\rm \delta UV$, over the observing time, $\rm T_{obs}$, is roughly: $\rm \delta UV \sim B_{uv} \times (T_{obs}/24hr)$, where $\rm B_{uv}$ is the baseline length (in meters or wavelengths). For the core baselines, $\rm B_{uv}$ is small, and essentially all the visibilities on a given baseline over the 3000s integration (300 visibilities for 10sec record length), fall into one or two uv-grid cells. For the outer baselines, the tracking speed is two orders of magnitude faster, and consecutive uv-points end up in individual grid-cells. In this case, UN vs NA weighting will make a big difference. 

The results for the image parameters are listed in Table 1. Columns 1 through 6 list, respectively: the array, the total time for the synthesis observation, the uv-weighting employed, the field size, the resulting synthesized beam FWHM (from Gaussian fitting to the dirty beam), and the image noise. 

Considering the first four rows, corresponding to the $50'$ field, the NA long and short observations give the same results, as expected, since each uv-data point is given the same weight, regardless of uv-grid cell size. For the UN weighting, the snapshot observation gives a very similar result as for NA weighting, because, again, the uv-cell size is small, such that each visibility is gridded into its own uv-cell, even for core baselines. On the other hand, the 3000s observation and UN weighting give a very different result: a much smaller synthesized beam, and higher noise. This is because the 300 visibilities along individual baseline time-tracks in the core are gridded into one or two cells, and hence get highly down-weighted, while the outer baselines track fast enough that, again, most visibilities are in cells of their own (see Figure 3). A curious consequence of the earth rotation synthesis observation is that gridded-weights, and image results, will depend on the record length as well. 

Next, consider rows 5 and 6 in Table 1. These correspond to instantaneous snapshot imaging, but synthesizing smaller field size, of $300"$ and $30"$. As the imaged field size gets smaller, the uv-grid cell size gets larger, and more and more baselines land in the same uv-cell, for the core baselines, but not for the long baselines (see Figure 2). In this case, UN vs. NA weighting will have a dramatic difference, with UN down-weighting the core baselines substantially, leading to smaller synthesized beams, and higher noise. 

Examples of the resulting synthesized beams are shown in the images in Figure 4, and as profiles through the beam in the East-West direction in Figure 5. Again, for the instantaneous snapshot images of a large field ($50'$), the UN and NA beams are very similar. For the smaller synthesized fields and UN weighting, the core baselines play less and less of a role, and the beam approaches that expected for the long baselines of the array. 

Also shown in Figure 5 is the Gaussian fit to the dirty beam in the UN, large field image, with a FWHM $= 2"$, which is, again, very similar to the NA beam. The UN wide field and NA dirty beams have much, much broader wings than a Gaussian, extending to $10"$ radius at the 20\% point. These wings have a major impact on image quality, even for point sources. Figure 6 shows the resulting image of a moderately bright point source ($7\sigma$), after deconvolution to the $s\sigma$ residual levels, for snapshot imaging of the UN wide field. The effect of the broad wing of the synthesized beam is   evident, with the source appearing 'fuzzy', extending to $3"$ radius at the $2\sigma$, or 30\%, level. Formally, a Gaussian fit to the source recovers the total flux to $\sim 20$\% accuracy, but the Gaussian fit implies a source extended on a scale comparable to the fitted Gaussian beam FWHM $\sim 2"$, even though the inserted model was a point source. 

The last four rows in Table 1 show a similar study for wide field snapshot vs. rotation synthesis imaging using the VLA-A configuration. The answer is similar to what was seen for the ngVLA, although the difference between UN and NA for the tracking images are less dramatic, since the ngVLA has the dense core, while the VLA does not. Also, the ngVLA has a higher spatial dynamic range (ratio of shortest to longest baseline), than a single configuration of the VLA. Hence, many more ngVLA short baselines get heavily down-weighted in the UN tracking image, vs. the snapshot image. We have also performed the same tests using real VLA data, with an observation at 8 GHz in A array of 3C286, and obtain results consistent with those shown in Table 1. 

\section{Spatially Correlated Noise}

Figure 7 shows the noise images for the UN snapshot image of a $5'$ field, and the NA image (again, results for the UN $50'$  snapshot image are much the same as NA). The character of the noise is very different for the two field sizes. The NA image shows large scale structure across the field, with characteristic size scales of order $30"$, corresponding to the spatial frequencies sampled by the core baselines. The UN $5'$ field image shows what appears more like spatial white noise. 

To quantify this effect, we Fourier transform the NA, and UN narrow field images, and examine the radial distribution of the power vs. baseline length, or uv-distance, for the Real and Imaginary parts of the visibilities. The results are shown in Figure 8. The UN, small field image shows what would be expected for a spatial distribution of white noise, ie. consistent with zero plus noise on all spatial scales for both the Real and Imaginary parts. The NA images show white noise in the Imaginary part, but the Real part shows substantial structure on baselines out to uv-distances $\rm \le 10k\lambda$, correspond to core baselines out to 1.25~km. 

\section{Summary}

We have explored instantaneous snapshot, narrow-band imaging with the ngVLA. Our primary conclusions are that:

\begin{itemize}
    \item The imaged field size plays a crucial role in determining the image parameters, even for pure UN weighting. A wide field implies a small uv-grid cell size, which can be small enough in the uv-plane such that each snapshot visibility is assigned to its own grid-cell, even for baselines in the dense core. In this case, UN weighting images approach NA weighting. Narrow fields lead to larger uv-grid cell sizes, and many different core baseline land in the same uv-cell, and hence get heavily down-weighted with UN weighting, leading to major differences between UN and NA images.
    
    \item Allowing for modest earth rotation synthesis changes this result substantially, since the uv-tracks for core baselines rotate much slower than outer baselines, such that many visibilities for a given core baseline land in single uv-cells, and hence get down-weighted with UN weighting. On long baselines, the visibilities track fast enough that individual records land in their own uv-cell.
    
    \item The resulting images for NA and UN, wide field, narrow band, instantaneous snapshot imaging are very similar, and display the imaging problems inherent in a core-dominated array, namely, very broad skirts that make point sources appear extended. 
    
    \item The spatial properties of the noise are also very different for the NA (and UN wide field snapshot), vs. UN narrow field images. The NA weighting leads to spatially correlated noise on scales comparable to the spatial frequencies sampled by the core baselines, in this case, out to $\sim 30"$. The UN narrower field images show spatial white noise.
    
\end{itemize}

The fundamental challenge in the case of instantaneous, narrow-band, wide field snapshot imaging is that UN weighted imaging approaches NA imaging, with all the deleterious effects of the wide skirts inherent in a NA weighted beam for a core dominated array, such as the ngVLA. If one is interested in a synthesized beam that is both narrow (width dictated by the longer baselines), and more Gaussian shaped, the only solution is to somehow down weight the core baselines.\footnote{While we have not seen similar calculations in the literature, we note that the SMA myriad users guide has the following statement: 'Surprisingly, making the field-of-view very large (bigger than the primary beam size), or very small (comparable to the synthesised beam), both cause uniform weighting to reduce to natural weighting,' with no explanation. The statement on small fields, comparable to the synthesized beam size, simply means, in the extreme, that all visibilities end up in one uv-cell. The large field statement is what we explore in detail in this memo. 
see: https://www.cfa.harvard.edu/sma/miriad/manuals/SMAuguide/smauserhtml/node107.html}

While the extreme applications considered herein, namely narrow band, wide field, high resolution, snapshot observations are likely to be uncommon in practice, the effect of uv-grid cell size will be apparent, to a lesser degree, in normal earth rotation and bandwidth synthesis observations, in particular for core-dominated arrays like the ngVLA. Modest Earth rotation synthesis down-weights the shorter baselines, as will bandwidth synthesis, when using UN weighting. Likewise, many wide field imaging algorithms use smaller facets to tile the sky in tangent planes. In this case, as well, the uv-grid cell sizes will be larger that for a full field, and hence the different weighting schemes will show different results.  However, in either case, the 'solution' to getting a cleaner beam is, in essence, no different than throwing out many of the core antennas, since they are heavily down-weighted in the imaging process using UN weighting. 

Overall, it is clear that, for science programs that require good imaging with resolutions afforded by the Plains baselines, and longer, most of the antennas in the core are not needed, and could be more usefully employed as a separate subarray for imaging diffuse cosmic structures.  The subarray solution has been explored in detail for the ngVLA in Rosero (2019, 2020). 


\vskip 0.2in
\noindent {\bf References}

\noindent Briggs, Schwab, Sramek 1999, ASP 180, 177 (SIRA II)

\noindent Carilli, C. 2017, ngVLA memo 16

\noindent Selina  et al. 2018, ASPC, 7, 15

\noindent Law et al. 2018a, ApJS, 236, 8

\noindent Law et al. 2018b, ASPC, 517, 773

\noindent Mooley et al. 2018, ApJ, 857, 143

\noindent Rosero 2019, ngVLA memo 55

\noindent Rosero 2020, ngVLA memo 76

\clearpage
\newpage

\begin{table}
\centering
\footnotesize
\caption{Effects of Weighting and Image Size}
\begin{tabular}{lccccc} 
\hline
Array & T$_{int}$ & UVWT & FoV & FWHM & RMS \\
\hline
~ & sec & ~ & arcsec & arcsec & mJy beam$^{-1}$ \\ 
\hline 
ngVLA & 0.1  &  NA &  3000 &  $2.14\times 2.08$ &  2.055 \\
ngVLA & 3000 &  NA &  3000 &  $2.14\times 2.08$ &  0.012 \\
ngVLA & 0.1  &  UN &  3000 &  $2.07\times 2.01$ &  2.103 \\
ngVLA & 3000 &  UN &  3000 &  $1.06\times 0.97$ &  0.019 \\
ngVLA & 0.1  &  UN &  300 &   $1.43\times 1.35$ &  2.641 \\
ngVLA & 0.1  &  UN &  30 &    $0.82\times 0.78$ &  6.008 \\
VLA-A & 0.1  &  NA &  3000 &  $0.92\times 0.86$ &  19.990 \\
VLA-A & 3000 &  NA &  3000 &  $0.92\times 0.86$ &  0.115 \\
VLA-A & 0.1  &  UN &  3000 &  $0.92\times 0.85$ &  20.080 \\
VLA-A & 3000 &  UN &  3000 &  $0.74\times 0.67$ &  0.134 \\
\hline
\vspace{0.1cm}
\end{tabular}
\end{table}

\begin{figure}
\centering 
\centerline{\includegraphics[scale=0.6]{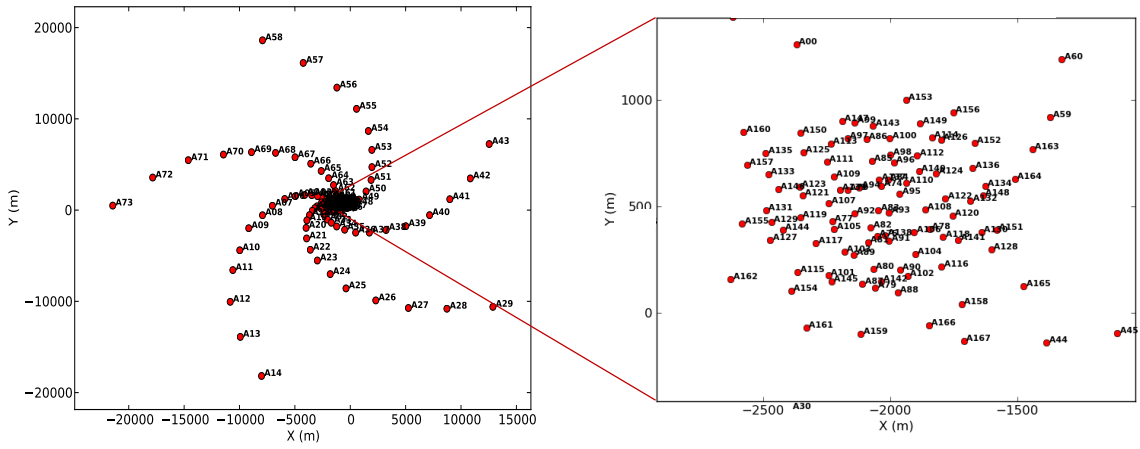}}
\caption{Left: The antenna distribution of Next Generation VLA Rev C Plains configuration, which includes a total of 168 antennas out to baselines of 40km. Right: Same, but just for the 94 core antennas.
}
\end{figure}  

\begin{figure}
\centering 
\centerline{\includegraphics[scale=0.65]{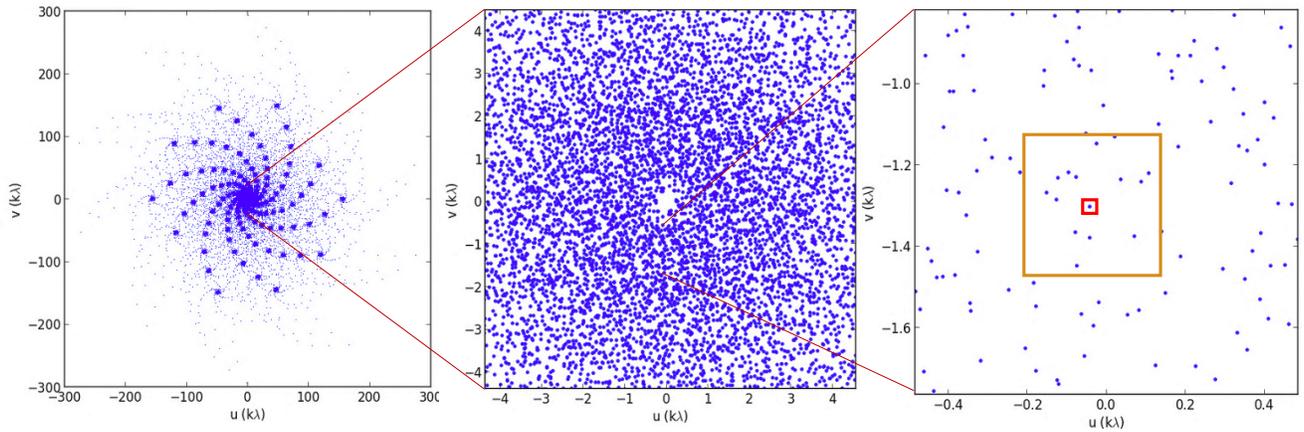}}
\caption{Left: The uv-plane coverage of the Next Generation VLA Rev C Plains configuration, for an instantaneous snapshot observation at 2.4 GHz or 0.125m. Center: The uv-plane coverage for the core baselines to 1.3km. Right: Expanded view of a dense region of the uv-plane corresponding to core baselines. The red box indicates the uv-cell size for a wide field image $= 50'$, or roughly the full primary beam to the 10\% power level, which equals 1/(2xImage size) = 34$\lambda$ = 4.3m. The orange box is the same, but for an imaged field $ = 5'$.
}
\end{figure}  

\begin{figure}
\centering  
\centerline{\includegraphics[scale=0.65]{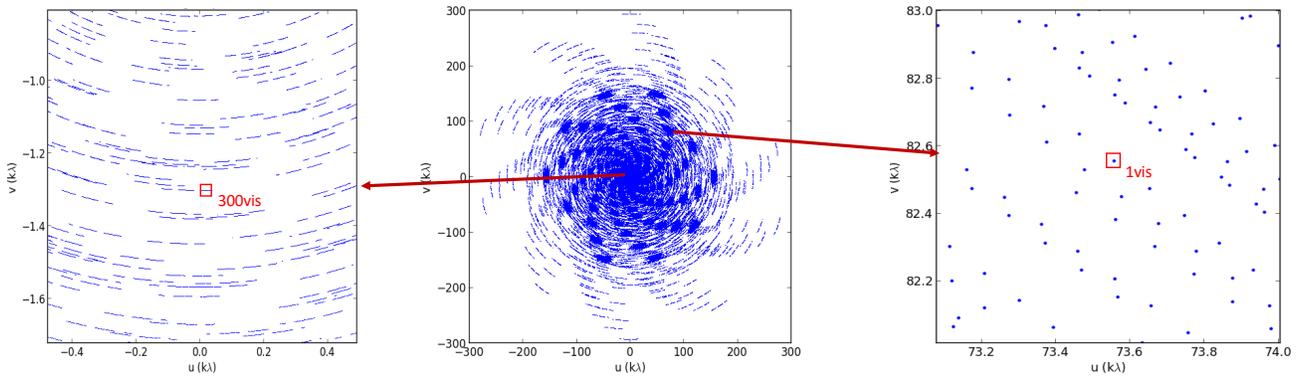}}
\caption{Center: The uv-plane coverage of the Next Generation VLA Rev C Plains configuration, for a 1 hour synthesis observation at 2.4 GHz or 0.125m. Left: Blow-up of a region of short baselines in the core ($\sim 160$m). Right: Blow-up of uv-tracks in a region of long baselines $\sim 14$km. The tracks for a given baseline run at an angle of about $45^o$ across the frame. The red box is the same size as in Figure 3, corresponding to the gridded cell size for a $50'$ image field of uvcell size = 34$\lambda$. In this simulation, the record length was 10sec. For the inner core baselines, each uv-cell contains 300 records, while on the long baselines, each uv-cell contains just a single visibility. 
}
\end{figure} 

\begin{figure}
\centering  
\centerline{\includegraphics[scale=0.65]{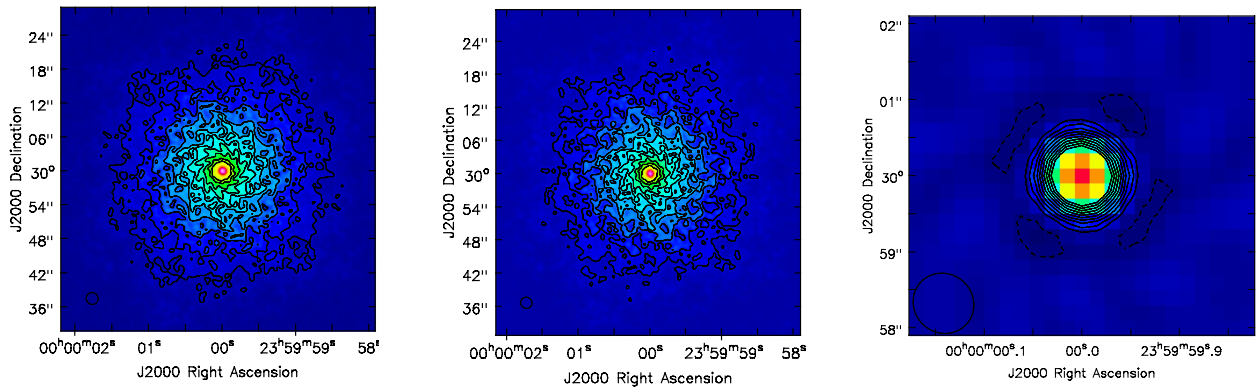}}
\caption{Left: The synthesized beam using pure NA weighting for a snapshot observation at 2.4 GHz for the Plains configuration. Center: same, but for UN weighting and a $50'$ field size. Right: Same, but for an imaged field size of only $0.5'$. In all cases, the contour levels are in steps of 5\%, starting at 5\%, up to 50\%, including two negative contours (dashed). 
}
\end{figure} 

\begin{figure}
\centering 
\includegraphics[scale=0.4]{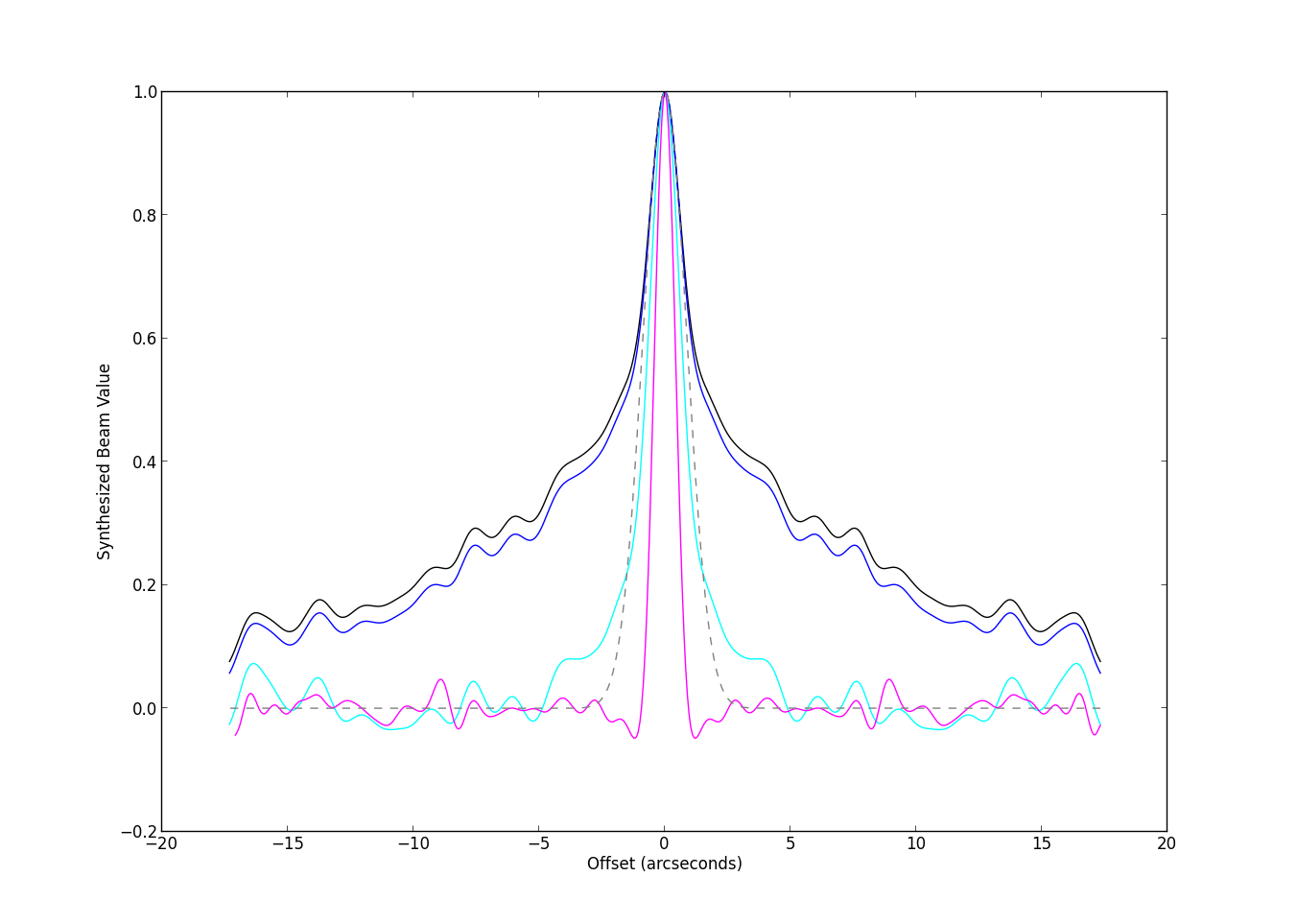}
\caption{Cuts through the synthesized beam for different weighting schemes and image sizes for a snapshot image. Black = NA weighting. Blue is UN weighting and a $50'$ field size. Cyan is UN and a $5'$ field size. Magenta is UN weighting with an $0.5'$ cell size. The grey dashed line is a Gaussian of FWHM = $2.04"$, corresponding to the fit to the UN weighting, $50'$ field synthesized beam. 
}
\end{figure}

\begin{figure}
\centering 
\centerline{\includegraphics[scale=0.7]{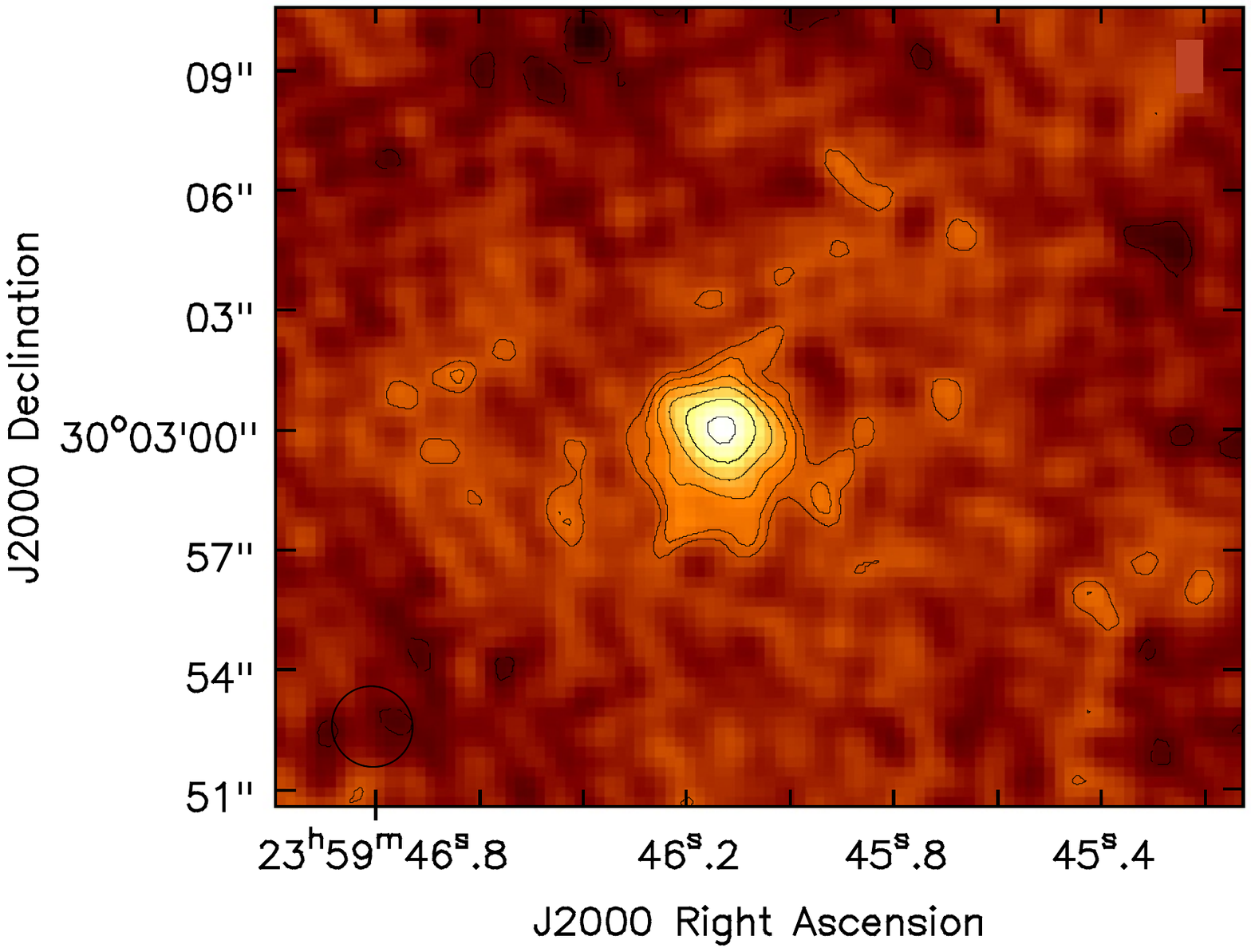}}
\caption{Image of a 7$\sigma$ point source in an image that has been deconvolved down to residual clean components of 2$\sigma$. The contour levels are a geometric progress in square root two, starting at 2$\sigma$. Note the broad skirt to $3"$ radius, even though the input model was a point source.
}
\end{figure}

\begin{figure}
\centering 
\centerline{\includegraphics[scale=0.65]{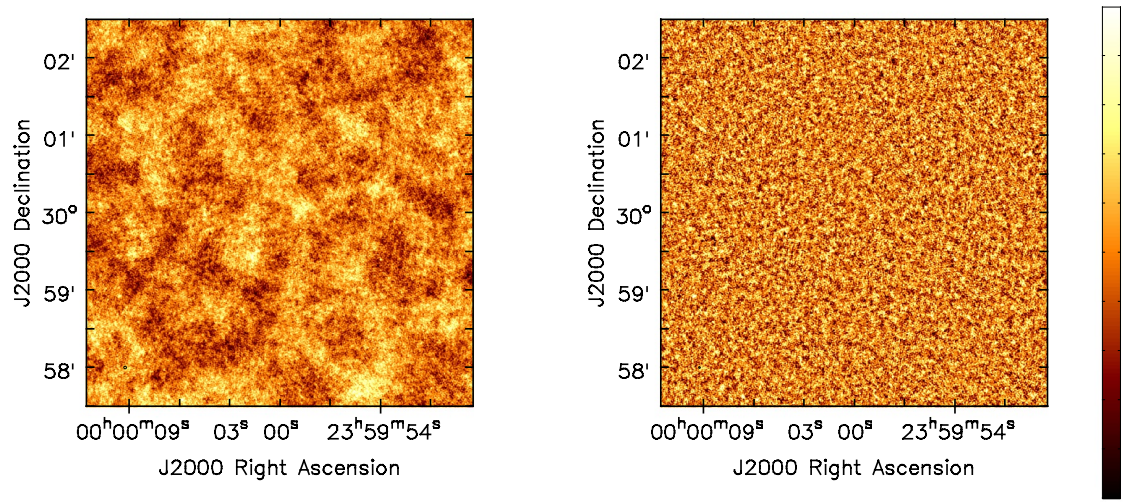}}
\caption{Left: Image of the noise for a NA weighted image for the 2.4 GHz snapshot observations with the ngVLA plains configuration. Right: Same, but for UN weighting an an image size of $5'$. The color scale ranges from $-0.4$ to 0.4 mJy beam$^{-1}$.
}
\end{figure}

\begin{figure}
\centering  
\centerline{\includegraphics[scale=0.7]{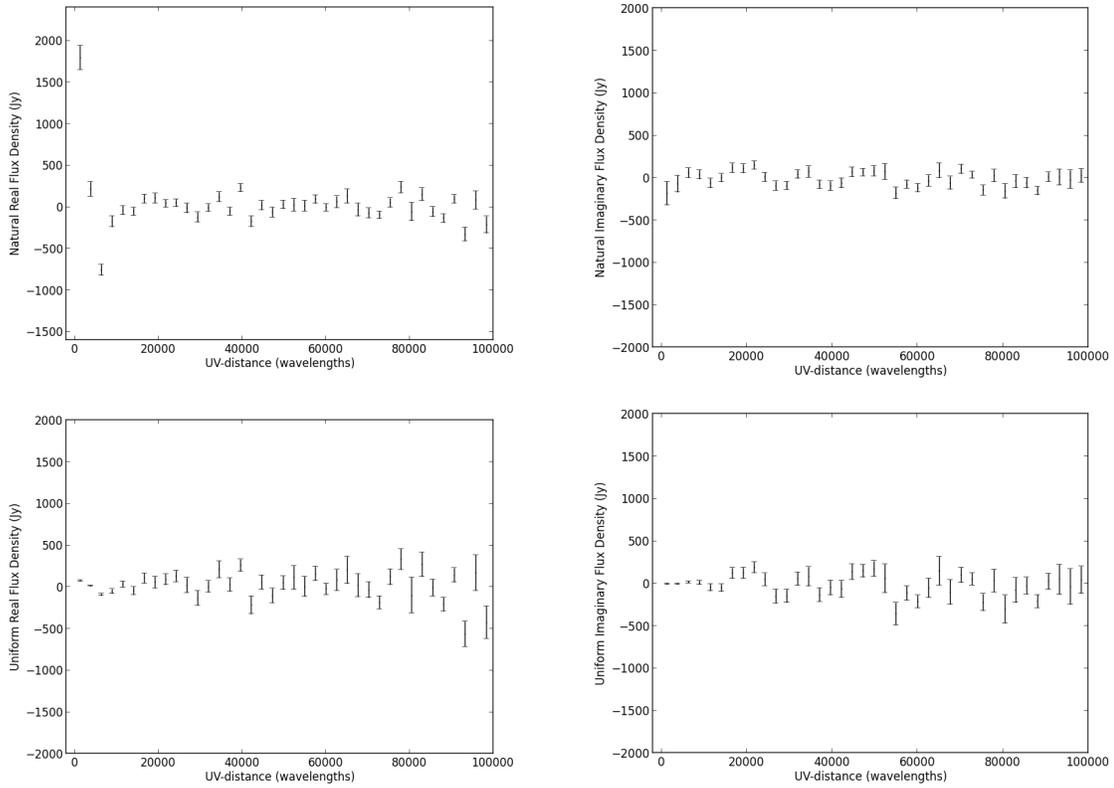}}
\caption{Top Left: UV-plot (amplitude vs. baseline length), for the Real part of the visibilities for visibilities made from a NA weighted image at 2.4 GHz with the ngVLA Plains configuration. Top Right: Same, but for the Imaginary part of the visibilities. Bottom Left: UV-plot (amplitude vs. baseline length), for the Real part of the visibilities for visibilities made from a UN weighted image at 2.4 GHz for a $5'$ field. Bottom Right: Same, but for the Imaginary part of the visibilities. Note that all the root power-spectra are consistent with zero, except for the NA Real, which shows distinct power at $\le 10$k$\lambda$ = 1.25km, ie. core baselines. This excess power can be seen in the spatially structured noise in Figure 7. 
}
\end{figure} 

\end{document}